\documentclass[prl,showpacs,superscriptaddress,twocolumn]{revtex4}

\usepackage{amsmath,bm}
\usepackage{graphicx}
\usepackage{epsfig}

\begin{document}

\newcommand{\vk}{{\vec k}}
\newcommand{\vK}{{\vec K}}
\newcommand{\vb}{{\vec b}}
\newcommand{{\vp}}{{\vec p}}
\newcommand{{\vq}}{{\vec q}}
\newcommand{\vQ}{{\vec Q}}
\newcommand{\vx}{{\vec x}}
\newcommand{\beq}{\begin{equation}}
\newcommand{\eeq}{\end{equation}}
\newcommand{\half}{{\textstyle \frac{1}{2}}}
\newcommand{\gton}{\stackrel{>}{\sim}}
\newcommand{\lton}{\mathrel{\lower.9ex \hbox{$\stackrel{\displaystyle<}{\sim}$}}}
\newcommand{\ee}{\end{equation}}
\newcommand{\ben}{\begin{enumerate}}
\newcommand{\een}{\end{enumerate}}
\newcommand{\bit}{\begin{itemize}}
\newcommand{\eit}{\end{itemize}}
\newcommand{\bc}{\begin{center}}
\newcommand{\ec}{\end{center}}
\newcommand{\bea}{\begin{eqnarray}}
\newcommand{\eea}{\end{eqnarray}}

\newcommand{\beqar}{\begin{eqnarray}}
\newcommand{\eeqar}[1]{\label{#1} \end{eqnarray}}
\newcommand{\pleft}{\stackrel{\leftarrow}{\partial}}
\newcommand{\pright}{\stackrel{\rightarrow}{\partial}}

\newcommand{\eq}[1]{Eq.~(\ref{#1})}
\newcommand{\fig}[1]{Fig.~\ref{#1}}
\newcommand{\eff}{ef\!f}
\newcommand{\alphas}{\alpha_s}

\renewcommand{\topfraction}{0.85}
\renewcommand{\textfraction}{0.1}
\renewcommand{\floatpagefraction}{0.75}

\title{ Production of  $\rho^{0}$ meson with large $p_{\rm T}$ at NLO in heavy-ion collisions}

\date{\today  \hspace{1ex}}
\author{Wei Dai}

\affiliation{School of Mathematics and Physics, China University of Geosciences (Wuhan), Wuhan 430074, China}

\author{Ben-Wei Zhang\footnote{bwzhang@mail.ccnu.edu.cn}}
\affiliation{Key Laboratory of Quark \& Lepton Physics (MOE) and Institute of Particle Physics,
 Central China Normal University, Wuhan 430079, China}

\author{Enke Wang}
\affiliation{Key Laboratory of Quark \& Lepton Physics (MOE) and Institute of Particle Physics,
 Central China Normal University, Wuhan 430079, China}

\begin{abstract}
Production of  large transverse momentum $\rho^{0}$ meson in high-energy nuclear collisions is investigated for the first time at the next-leading-order in the QCD improved parton model.  The $\rho^0$ fragmentation functions (FFs) in vacuum at any scale $Q$ are obtained, by evolving a newly developed initial parametrization of $\rho^0$ FFs  at a scale $\rm Q_{0}^2=1.5\ GeV^2$ from a broken SU(3) model through NLO DGLAP equations. The numerical simulations of $p_{\rm T}$ spectra of $\rho^{0}$ meson in the elementary $\rm p+p$ collisions at NLO give a decent description of STAR $\rm p+p$ data. In $\rm A+A$ reactions the jet quenching effect is taken into account with the higher-twist approach by the medium-modified parton FFs due to gluon radiation in the quark-gluon plasma, whose space-time evolution is described by a (3+1D) hydrodynamical model.  The nuclear modification factors for $\rho^{0}$ meson and its double ratio with $\pi^\pm$ nuclear modification in central $\rm Au+Au$ collisions at the RHIC are calculated and found to be in good agreement with STAR measurement. Predictions of $\rho^{0}$ nuclear modification and the yield ratio $\rho^0/\pi^0$ in central Pb+Pb at the LHC are also presented. It is shown that the ratio $\rho^0/\pi^0$ in central Pb+Pb will approach to that in p+p reactions when $p_{\rm T}>12$~GeV.
\end{abstract}


\maketitle

A new state of matter of deconfined quarks and gluons, the so-called quark-gluon plasma (QGP), is expected to be created in heavy ion collisions (HIC) at very high colliding energies. To study the creation and properties of the QGP, the jet quenching has been proposed, which states that when an energetic parton traveling through the hot/dense QCD medium, a substantial fraction of its energy should be losted and could in turn be used to obtain the temperature and density information of the QGP~\cite{Wang:1991xy,Gyulassy:2003mc}. Even though rapid developments of experiments and theories on new jet quenching observables, such as di-hadron~\cite{Aamodt:2011vg, Adler:2002tq}, photon triggered  hadron~\cite{Adare:2009vd,Abelev:2009gu} and full jet observable~\cite{Vitev:2008rz,Vitev:2009rd,Dai:2012am,Aad:2010bu,Chatrchyan:2011sx,Kang:2014xsa}, have emerged in the last decade, the suppression of inclusive hadron production, as the most intensively studied observable on jet quenching, is still indispensable to unravel the properties of the QCD medium. 
Recently, by comparing the theoretical calculation with the measurements of the production spectra and its suppression of $\pi$ mesons which are the most commonly observed hadrons, the jet transport coefficient $\hat{q}$ has been extracted to characterize the local properties of the QCD medium probed by the energetic parton jets~\cite{Burke:2013yra}.  The higher twist multiple scattering of the jet quenching incorporated with perturbative quantum chromodynamics (pQCD) improved parton model has been developed and successfully described the $\pi^0$ and $\eta$ productions and their suppressions in $\rm A+A$ collisions~\cite{Chen:2010te, Chen:2011vt, Dai:2015dxa,Dai:2016zjy}.

The study of the identified hadron spectra at high $p_{\rm T}$ other than $\pi^0$ and $\eta$ in HIC can further constrain and cast insight into the hadron suppression pattern. Whereas a relatively large amount of data on the yields of identified hadrons at large $p_{\rm T}$ has been accumulated at the RHIC and the LHC~\cite{Agakishiev:2011dc,Adare:2010pt,Bala:2016hlf}, there are still very few theoretical studies of hadrons with different types.   An interesting type of identified hadrons with available data is $\rho^0$ meson, which is heavier than $\pi^{0}$ and $\eta$, and also consists of the similar constituent quarks. We notice that even the theoretical calculations of the $\rho^{0}$ productions in p+p collisions with large $p_{\rm T}$ at both the RHIC and the LHC are absent due to the lack of knowledge of parton fragmentation functions (FFs) for $\rho_0$ in vacuum. In a previous study~\cite{ Dai:2015dxa} we have paved the way to understand identified hadron suppression pattern by calculating the productions of $\eta$ meson and investigating the hadron yield ratios~\cite{Dai:2015dxa}.
In this manuscript, we extend this study to $\rho^{0}$ meson productions and the yield ratios of $\rho^{0}$ and $\pi$ in A+A collisions at the RHIC and the LHC. It is of great interest to see how the alteration of the jet chemistry brought by the jet quenching will eventually affect the $\rho^{0}$ production spectrum and the ratio of hadron yields~\cite{Liu:2006sf,Brodsky:2008qp,Chen:2008vha}.

In this paper, firstly we employ
a newly developed initial parametrization of $\rho^0$ FFs  in vacuum at a starting scale $\rm Q_{0}^2=1.5\ GeV^2$, which is provided by the $SU(3)$ model of FFs of vector mesons~\cite{Saveetha:2013jda,Indumathi:2011vn}.  By evolving them through DGLAP evolution equations at NLO~\cite{Hirai:2011si}, we obtain parton FFs of $\rho^{0}$ meson at any hard scale $Q$. The theoretical results of $\rho^{0}$ productions in $\rm p+p$ collisions are provided up to the next-to-leading order(NLO) in pQCD improved parton model, and we find that they describe the experimental data rather well. Then we study $\rho^{0}$ production in $\rm A+A$ collisions at both RHIC and LHC by including parton energy loss in the hot/dense QCD medium in the framework of higher twist approach of jet quenching~\cite{Guo:2000nz,Zhang:2003yn, Zhang:2003wk}. In this approach, the energy loss due to the multiple scattering suffered by an energetic parton traversing the medium are taken into account by twist-4 processes, and the vacuum fragmentation functions are modified effectively in high-energy nuclear collisions. Therefore, we can compute numerically for the first time $\rho^{0}$ meson yields in $\rm A+A$ collisions. We give a description of $\rho^{0}$ nuclear modification factor $R_{AA}(\rho^0)$ at large $p_{\rm T}$ in $\rm Au+Au$ collisions at the RHIC to confront against the experimental data by STAR Collaboration, and $R_{AA}(\rho^0)$ in $\rm Pb+Pb$ collisions at the LHC to give a theoretical prediction. The double ratio of $R_{AA}(\rho^0)/R_{AA}(\pi^\pm)$ is calculated and found to be in good agreement with the experimental data. Lastly we explore the features of the $\rho^0/\pi^0$ ratios in both p+p and A+A collisions.

In NLO pQCD calculation, the single hadron production can be factorized as the convolution of elementary partonic scattering cross sections up to $\alpha^3$, parton distribution functions (PDFs) inside the incoming particles and parton FFs to the final state hadrons~\cite{Kidonakis:2000gi}. We can express the formula symbolically as:
\begin{eqnarray}
\frac{1}{p_{T}}\frac{d\sigma_{h}}{dp_{T}}=\int F_{q}(\frac{p_{T}}{z_{h}})\cdot D_{q\to h}(z_{h}, p_{T})\frac{dz_{h}}{z_{h}^2} \nonumber  \\
+ \int F_{g}(\frac{p_{T}}{z_{h}})\cdot D_{g\to h}(z_{h}, p_{T})\frac{dz_{h}}{z_{h}^2}  \,\,\, .
\label{eq:ptspec}
\end{eqnarray}
The above equation implies that the hadron yield in $\rm p+p$ collision will be determined by two factors: the initial (parton-)jet spectrum $F_{q,g}(p_T)$ and the parton fragmentation functions $D_{q,g\to h}(z_{h}, p_{T})$.  In the following calculations, we utilize CTEQ6M parametrization for proton PDFs ~\cite{Lai:1999wy}, which has been convoluted with the elementary partonic scattering cross sections up to $\alpha^3$ to obtain $F_{q,g}(\frac{p_{T}}{z_{h}})$. Here $D_{q,g\to h}(z_{h}, p_{T})$ represents the vacuum parton FFs, which denote the possibilities of scattered quark or gluon fragmenting into hadron $h$ with momentum fraction $z_h$. They can be given by corresponding parametrization for different final-state hadrons. So potentially, we could predict all the identified hadron productions in $\rm p+p$ collision as long as the fragmentation functions are available. Note that the factorization scale, renormalization scale and fragmentation scale are usually chosen to be the same and proportional to $\rm p_{\rm T}$ of the leading hadron in the final-state.

To accurately determine the $\rm p+p$ reference,  parton FFs in vacuum as a non-perturbative input, should be available. So far it is still impossible to derive parton FFs from the first-principle of QCD and a common practice is to make phenomenological parametrizations by comparing perturbative QCD calculations with the data. Unlike $\pi$ and charged hadrons, until now there are very few satisfatory parametrizations of parton FFs for the vector mesons due to the paucity of the relevant data. Fortunately, a broken $SU(3)$ model is recently proposed to provide a systematic description of the vector mesons production~\cite{Saveetha:2013jda,Indumathi:2011vn}. To reduced the complexity of the meson octet fragmentation functions, the $SU(3)$ flavor symmetry is introduced with a symmetry breaking parameter. In addition, isospin and charge conjugation invariance of the vector mesons $\rho(\rho^+,\rho^-,\rho^{0})$ are assumed to further reduce independent unknown quark FFs into functions named valence(V) and sea($\gamma$). The inputs of valence $V(x, Q_0^2)$, sea $\gamma(x, Q_0^2)$ and gluon $D_g(x,Q_0^2)$ FFs are parameterized into a standard polynomial at a starting low energy scale of $Q_0^2=1.5$~$\rm GeV^2$ such as:
\begin{eqnarray}
F_i(x)=a_ix^{b_i}(1-x)^{c_i}(1+d_ix+e_ix^2)
\end{eqnarray}
These parameters are systematically fixed by fitting the cross section at NLO with the measurements of LEP($\rho$,$\omega$) and SLD($\phi$,$K^\star$) at $\sqrt{s}=91.2$~GeV. In Ref.~\cite{Saveetha:2013jda,Indumathi:2011vn} the parameters of $\rho^{0}$ FFs in vacuum at $Q^2=1.5$~$\rm GeV^2$ are listed and we obtain $\rho^{0}$ FFs at any hard scale $D_{q,g}(x,Q^2)$ $Q>2$~GeV by evolving them through DGLAP evolution equations at NLO with the compute code invented in Ref.~\cite{Hirai:2011si}, then these $\rho_0$ FFs $D_{q,g}(x,Q^2)$ are used in our numerical simulations.

\begin{figure}[!h]
\begin{center}
\hspace*{-0.1in}
\vspace*{-0.1in}
\includegraphics[width=1.7in,height=1.7in,angle=0]{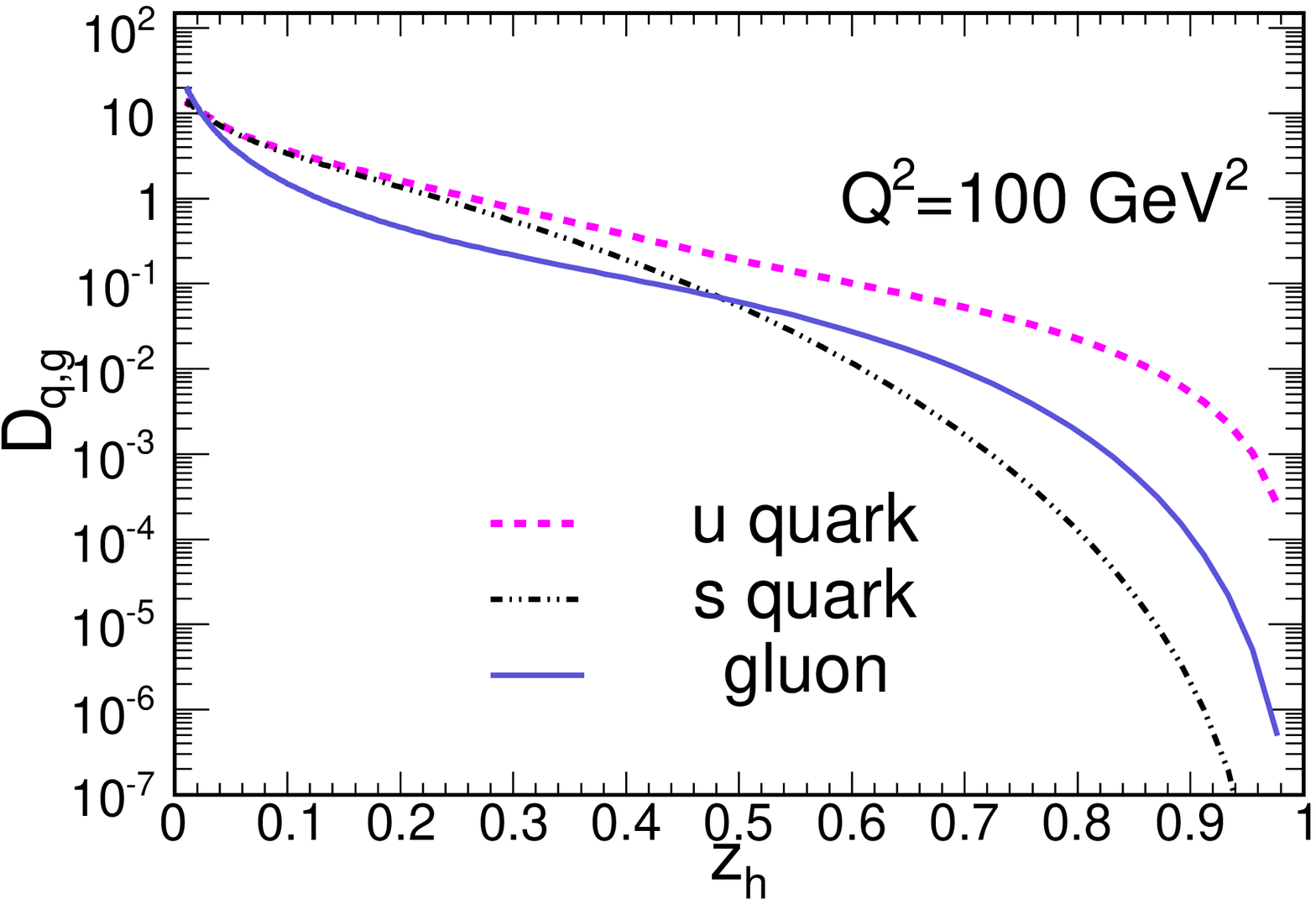}
\includegraphics[width=1.7in,height=1.7in,angle=0]{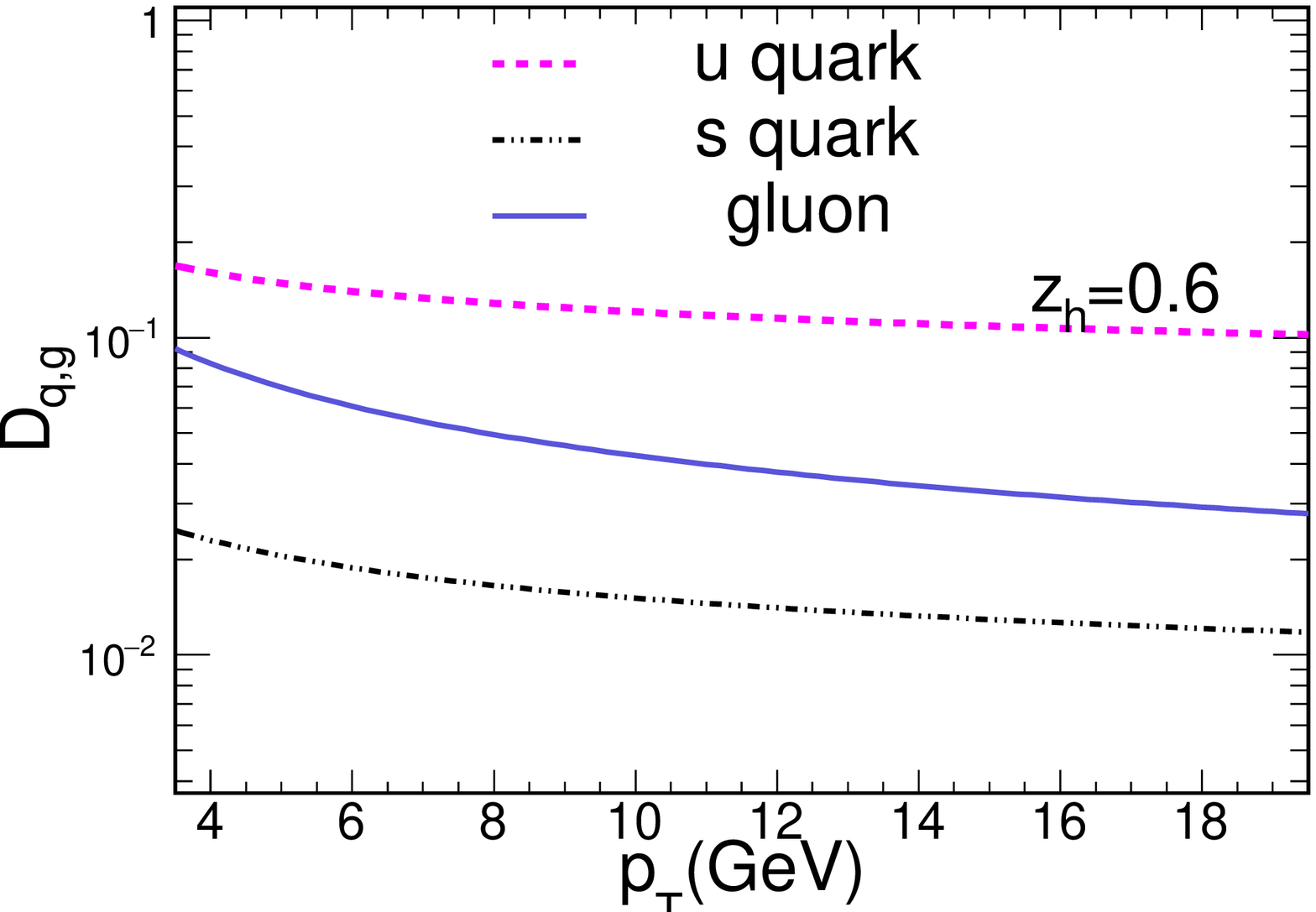}
\hspace*{-0.1in}
\vspace*{0.0in}
\caption{ Left: parton FFs as functions of $z_h$ at fixed scale $Q^2=100$~$\rm GeV^2$; 
Right: parton FFs as functions of $p_{\rm T}$ at fixed $z_h=0.6$.}
\label{fig:fragfunc}
\end{center}
\end{figure}

 We have plotted the parton FFs as functions of fragmenting fraction $z_h$ in the left panel of Fig.~\ref{fig:fragfunc} at fixed scale of $Q^2=100$~$\rm GeV^2$, and also the parton FFs as functions of final state $\rm p_{\rm T}$ at fixed fragmenting fraction $z_h=0.6$ in the right panel of Fig.~\ref{fig:fragfunc}. It is observed that at fixed scale $\rho^{0}$ FFs decrease with $z_h$, and FF of  up quark is much larger than that of strange quark, especially at large $z$ region. At a typical value with $z_h=0.6$, we notice that $\rho^{0}$ FFs show a rather weak $p_{\rm T}$ dependence.

\hspace{0.7in}
\hspace*{-1.5in}

\begin{figure}[!h]
\begin{center}
\hspace*{-0.1in}
\includegraphics[width=3.4in,height=2.8in,angle=0]{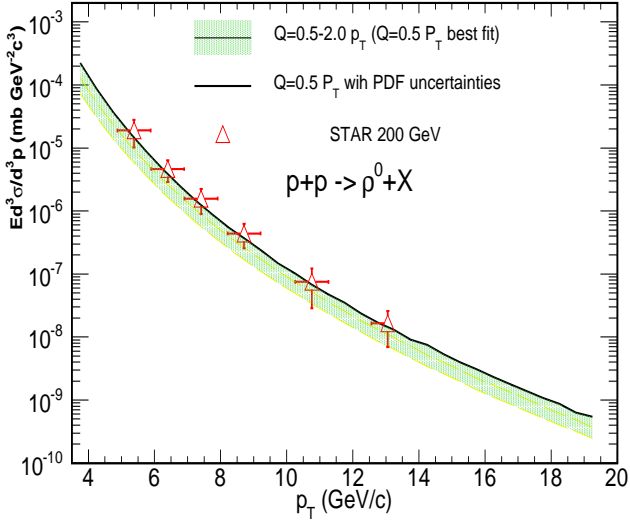}
\hspace*{-0.1in}
\vspace*{-0.2in}
\caption{ Numerical calculation of the $\rho^{0}$ production in $\rm p+p$ collisions at RHIC $200$~GeV comparing with STAR~\cite{Agakishiev:2011dc} data.}
\label{fig:illustrhopp}
\end{center}
\end{figure}

The existence of the $\rho^0$ meson FFs at NLO allows us to calculate the inclusive vector meson productions as a function of the final state hadron $p_{\rm T}$ in pQCD at the accuracy of NLO.  Fig.~\ref{fig:illustrhopp} shows the confrontation of the theoretical calculation with the STAR data~\cite{Agakishiev:2011dc}. We see the results at the scale $Q=0.5$~$p_{\rm T}$ agree well with the data of $\rho^0$ yield. In the following calculations we will fix  $Q=0.5$~$p_{\rm T}$ to provide a good p+p baseline.

\hspace{0.7in}
\hspace*{-0.5in}

A hot and dense QCD matter is created shortly after the high energy central nucleus-nucleus collisions. Before a fast parton fragmented into identified hadrons in the vacuum, it should suffer energy loss due to multiple scattering with other partons in QCD medium. In higher twist approach, the multiple scattering is described by twist-4 processes of hard scattering and will lead to effective medium-modification of the vacuum FFs~\cite{Guo:2000nz, Zhang:2003yn, Zhang:2003wk, Chen:2010te,Chen:2011vt, Dai:2015dxa, Dai:2017piq}:
\begin{eqnarray}
\tilde{D}_{q}^{h}(z_h,Q^2) &=&
D_{q}^{h}(z_h,Q^2)+\frac{\alpha_s(Q^2)}{2\pi}
\int_0^{Q^2}\frac{d\ell_T^2}{\ell_T^2} \nonumber\\
&&\hspace{-0.7in}\times \int_{z_h}^{1}\frac{dz}{z} \left[ \Delta\gamma_{q\rightarrow qg}(z,x,x_L,\ell_T^2)D_{q}^h(\frac{z_h}{z}, Q^2)\right.
\nonumber\\
&&\hspace{-0.2 in}+ \left. \Delta\gamma_{q\rightarrow
gq}(z,x,x_L,\ell_T^2)D_{g}^h(\frac{z_h}{z}, Q^2) \right] ,
\label{eq:mo-fragment}
\end{eqnarray}
where $\Delta\gamma_{q\rightarrow qg}(z,x,x_L,\ell_T^2)$ and $\Delta\gamma_{q\rightarrow gq}(z,x,x_L,\ell_T^2)=\Delta\gamma_{q \rightarrow qg}(1-z,x,x_L,\ell_T^2)$ are the medium modified splitting functions~\cite{Guo:2000nz, Zhang:2003yn, Zhang:2003wk}. Though the medium-modified FFs include a contribution from gluon radiation in the QCD medium, they obey QCD evolution equations similar to  the DGLAP equations for FFs in vacuum. In this formalism, we convolute the medium-induced kernel $\Delta\gamma_{q\rightarrow qg}(z,x,x_L,\ell_T^2)$ and $\Delta\gamma_{q\rightarrow gq}$ (instead of those vacuum splitting functions) with the (DGLAP) evolved FFs at scale $Q^2$. We average the above medium modified fragmentation functions over the initial production position and jet propagation direction, scaled by the number of binary nucleon-nucleon collisions at the impact parameter $b$ in $\rm A+A$ collisions to replace the vacuum fragmentation functions in Eq.~(\ref{eq:ptspec}). In the medium modified splitting functions $\Delta\gamma_{q\rightarrow qg,gq}$, we can extract the dependency of the properties of the medium into the jet transport parameter $\hat{q}$ which defined as the average squared transverse momentum broadening per unit length. In the higher-twist approach, the jet transport parameter $\hat{q}$ is related to the gluon distribution density of the medium.
 Phenomenologically the jet transport parameter can be assumed to be proportional to the local parton density in the QGP phase and also to the hadron density in the hadronic gas phase~~\cite{Chen:2010te}:
\begin{equation}
\label{q-hat-qgph}
\hat{q} (\tau,r)= \left[\hat{q}_0\frac{\rho_{QGP}(\tau,r)}{\rho_{QGP}(\tau_{0},0)}
  (1-f) + \hat q_{h}(\tau,r) f \right]\cdot \frac{p^\mu u_\mu}{p_0}\,,
\end{equation}
$\rho_{QGP}$ is the parton (quarks and gluon) density in an ideal gas at a given temperature,
$f(\tau,r)$ is the fraction of the hadronic phase as a function of
space and time, $\hat q_{0}$ is the jet transport
parameter at the center of the bulk medium in the QGP phase at the
initial time $\tau_{0}$, $p^\mu$ is the four momentum of the jet and $u^\mu$ is the
four flow velocity in the collision frame.

The space-time evolution of the QCD medium is given by a full three-dimensional (3+1D) ideal hydrodynamics description~\cite{Hirano2001,HT2002}. Parton density, temperature, fraction of the hadronic phase and the four flow velocity at every time-space points are provided by the hydro dynamical model.  The only free parameter is $\hat{q}_0\tau_0$, the product of initial value of the jet transport parameter $\hat{q}_0$ and the time $\tau_0$ when the QCD medium is initially formed. This parameter controls the strength of jet-medium interaction,  and the amount of the energy loss of the energetic jets. In the calculations, we use the values of $\hat{q}_0\tau_0$ extracted in the previous studies~\cite{Chen:2010te, Chen:2011vt, Dai:2015dxa}, which give very nice descriptions of single $\pi^0$ and $\eta$ productions in HIC. Moreover, we have used the EPS09 parametrization sets of nuclear PDFs  $f_{a/A}(x_a,\mu^2)$ to consider the initial-state cold nuclear matter effects~\cite{Eskola:2009uj}.

\hspace{0.7in}
\begin{figure}[!h]
\begin{center}
\hspace*{-0.1in}
\vspace*{-0.1in}
\includegraphics[width=3.0in,height=3.2in,angle=0]{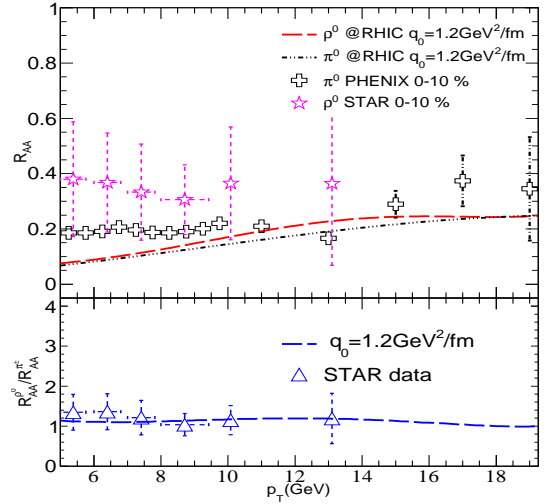}
\vspace*{-0.4in}
\caption{Top panel: Numerical calculation of the $\rho^{0}$ and $\pi^0$ production suppression factors in $0-10\%$ $\rm Au+Au$ collisions at RHIC $200$~GeV at NLO as functions of $p_{\rm T}$, comparing with STAR~\cite{Agakishiev:2011dc} and PHENIX~\cite{Adler:2003qi} data; Bottom panel: double ratio calculation of $R_{\rm AA}^{\rho^0}/R_{\rm AA}^{\pi^\pm}$ both at NLO, also comparing with STAR data.
}
\label{fig:illustrhorhic}
\end{center}
\end{figure}
\hspace*{-0.5in}

\hspace{0.7in}
\begin{figure}[!h]
\begin{center}
\hspace*{-0.1in}
\vspace*{-0.1in}
\includegraphics[width=3.0in,height=2.2in,angle=0]{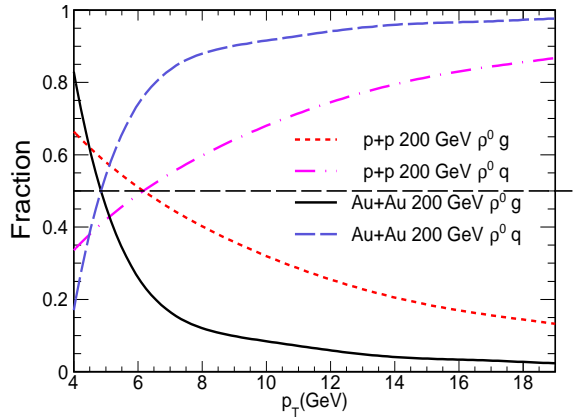}
\vspace*{-0.0in}
\caption{Gluon and quark contribution fraction of the total yield both in p+p and Au+Au at RHIC
}
\label{fig:rhofracrhic}
\end{center}
\end{figure}


Now we are ready to calculate the single $\rho^{0}$ productions in heavy ion collisions up to the NLO. The nuclear modification factor $R_{\rm AA}$ as a function of  $p_{\rm T}$ is  calculated to demonstrate the suppression of the production spectrum in $\rm A+A$ collisions relative to that in $\rm p+p$ collision:
\begin{eqnarray}
R_{AB}(b)=\frac{d\sigma_{AB}^h/dyd^2p_T}{N_{bin}^{AB}(b)d\sigma_{pp}^h/dyd^2p_T}
\label{eq:eloss}
\end{eqnarray}

In the $0-10\%$ most central $\rm Au+Au$ collisions at RHIC $200$~GeV, we calculate $\rho^{0}$ productions at typical values of $\hat q_{0}=1.2$~GeV$^2$/fm and
$\tau_{0}=0.6$~fm at the RHIC~\cite{Dai:2015dxa}.  The theoretical calculation can explain the data of $\rho^{0}$ meson at large $p_{\rm T}$ region (see the top panel of Fig.~\ref{fig:illustrhorhic}). The theoretical calculation and the experimental data of the $\pi^0$ nuclear suppression factor are also presented for comparison. We note that the nuclear suppression factor of $\rho^0$ is similar to the one of $\pi^{0}$, as demonstrated by the double ratio $R_{\rm AA}^{\rho^0}/R_{\rm AA}^{\pi^\pm}$ in the bottom panel of Fig.~\ref{fig:illustrhorhic}, which is around 1 calculated at the NLO accuracy. We also find that the theoretical curve undershoots the experimental data of $R_{\rm AA}$ same as the case in $\pi^0$, and the uncertainty caused by this undershooting will be cancelled out to a large extent when we discussing the double ratio of $\rho^0$ and charged $\pi$. Here $\pi^{\pm}$ FFs in vacuum are given by AKK08~\cite{Albino:2008fy}.

To understand better the nature of the suppression pattern of $\rho^0$, we calculate the gluon (quark) contribution fraction of the total yield both in $\rm p+p$ and $\rm Au+Au$ collisions in Fig.~\ref{fig:rhofracrhic}. It is similar to $\eta$ and $\pi^0$ productions which demonstrate the domination of the quark fragmentation process contribution at high $p_{\rm T}$ region either in $\rm p+p$ or in 
$\rm A+A$ collisions , and the jet quenching effect may suppress the gluon fragmenting contribution but enhance the quark contribution. Therefore the crossing point where the fractional contributions of quark and gluon fragmentation are equal, will move toward lower $p_{\rm T}$ in $\rm Au+Au$ collision, as one observes in Fig.~\ref{fig:rhofracrhic}.

\hspace{0.1in}
\vspace*{-0.1in}
\begin{figure}[!h]
\begin{center}
\hspace*{-0.1in}

\includegraphics[width=3.6in,height=4.4in,angle=0]{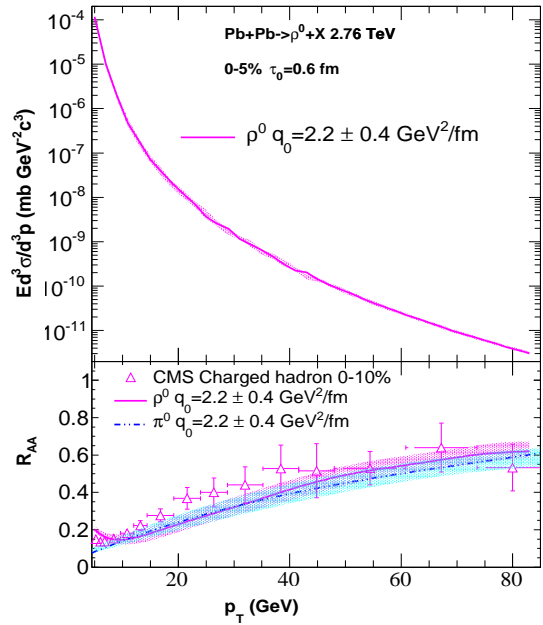}
\hspace*{-0.1in}
\vspace*{-1.2in}
\caption{ Numerical calculation of the $\rho^{0}$ production in $0-10\%$ $\rm Pb+Pb$ collisions at LHC $2.76$~GeV in the top panel; theoretical calculation results of nuclear suppression factor of $\rho^{0}$ and $\pi^0$ are compared with the experimental data of charged hadron~\cite{Aamodt:2010jd} in $0-10\%$ $\rm Pb+Pb$ collisions at LHC $2.76$~GeV in the bottom panel.}
\label{fig:illustrholhc}
\end{center}
\end{figure}
\hspace*{-0.5in}

We also predict the $\rho^{0}$ production in the $0-10\%$ most central $\rm Pb+Pb$ collisions at the LHC with $\sqrt{s_{NN}}=2.76$~TeV in the top panel of  Fig.~\ref{fig:illustrholhc}. The values of the $\hat q_{0}$ are set to be the same as the typical values which have been used to describe production suppression of both single $\pi^0$ and $\eta$ mesons at the LHC~\cite{Chen:2010te, Chen:2011vt, Dai:2015dxa}. We can see that, with the increase of $p_{\rm T}$, the nuclear modification factor of $\rho^0$ meson goes up slowly. In the calculation, best fit to the PHENIX data on $\pi^0$ nuclear suppression factor as a function of $p_{\rm T}$ in $0-5\%$ Au+Au collisions at $\sqrt{s}=200$~$  \rm GeV$ gives $\hat{q}_0=1.20\pm0.30$~$\rm GeV^2/fm$. Similarly, the best fit to the CMS data on charged hadron nuclear suppression factor in $0-5\%$ Pb+Pb collisions at $\sqrt{s}=2.76$~$  \rm TeV$ as a function of $p_{\rm T}$ would gives $\hat{q}_0=2.2\pm0.4 ~\rm GeV^2/fm$ at $\tau_0=0.6 ~\rm fm/c$~\cite{Burke:2013yra}.  The same values of $\hat{q}_0\tau_0$ have been employed to give a very nice description of $\rho^0$ productions in LHC shown in the bottom panel of  Fig.~\ref{fig:illustrholhc}.

\begin{figure}[!h]
\begin{center}
\hspace*{-1.1in}
\vspace*{-0.1in}
\includegraphics[width=3.0in,height=2.4in,angle=0]{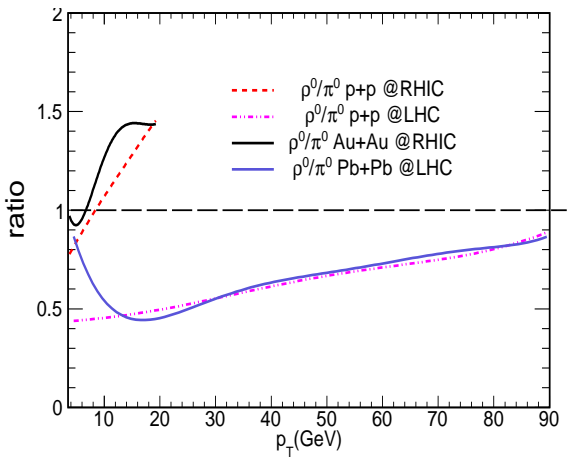}
\hspace*{-0.9in}
\vspace*{-.0in}
\caption{ $\rho^0/\pi^0$ production ratio as a function of final state $p_{\rm T}$ calculated both in p+p and A+A collisions at RHIC and LHC }
\label{fig:ratio}
\end{center}
\end{figure}

To compare the different trends of $\pi^0$ and $\rho^0$ spectra, we plot the ratio $\rho^0/\pi^0$ as a function of the transverse momentum $p_{\rm T}$ in Fig.~\ref{fig:ratio}.
As we have mentioned that in the study the $\pi^0$ FFs are given by 
AKK08~\cite{Albino:2008fy}. We note that even the validity of the $\pi^0$ (charged hadron) FFs had been challenged by the over-predicting of its production in the LHC and Tevatron due ot the too-hard gluon-to-hadron FFs in the parameterizations~\cite{dEnterria:2013sgr}.  A recent attempt to address the problem and a global refit is performed in Ref.~\cite{deFlorian:2014xna}.  The uncertainty brought in by the usage of AKK08 fortunately do not affect the results of the nuclear modification factor $R_{\rm AA}$ much due to the cancellation when taking the ratio of A+A production to p+p reference. Therefore one expects that the extraction of jet transport parameter $\hat{q}_0$ from the comparison between theoretical calculated $R_{\rm AA}$ and the experimental data will not be affected much by such FFs uncertainties. In the studies of particle ratio, $\pi^0$ fragmentation function and its jet chemistry are used as reference to understand other mesons such as $\eta$, its FFs uncertainties certainly will be expected to affect particle ratios like $\eta/\pi^0$.  However, since light mesons such as $\pi^0$ and $\eta$ are dominated by quark fragmenting contribution, such effect is therefore minimized.

Fig.~\ref{fig:ratio} illustrates that the ratio $\rho^0/\pi^0$  increases with the $\rm p_{\rm T}$ in $\rm p+p$ collision at the RHIC energy and LHC.  Though the jet quenching effect may alter the ratio a little bit in $\rm A+A$ at lower $\rm p_T$, as $\rm p_T$ becomes larger, the ratio in $\rm A+A$ comes very close that in $\rm p+p$, especially at the LHC with higher $p_{\rm T}$.   We note flat curves are observed in $\eta/\pi^0$ ratios as functions of $p_{\rm T}$ at both the RHIC and the LHC, whereas a increasing $\rho^0/\pi^0$ with respect to $p_{\rm T}$ are shown in  Fig.~\ref{fig:ratio}. The $\rho^0/\pi^0$ ratio in the RHIC demonstrates a more rapidly increasing behavior with respect to $p_{\rm T}$.  It is realized that the flat particle ratio dependence of $p_{\rm T}$ is therefore not a universal trend, and the shape of the particle ratio depends on the relative slope of their spectra 
in p+p  , different flavor contributions to FFs as well as flavor dependence of parton energy loss in the QGP.

\hspace{0.0in}
\hspace*{-0.5in}

We note that at high $p_{\rm T}$ region, the productions of both $\rho^0$ and $\pi^0$ are dominated by quark contribution (for example, see Fig.~\ref{fig:rhofracrhic}). If at high $p_{\rm T}$, quark FFs of $\rho^0$ and $\pi^0$ have a relatively weak dependence on $z_h$ and $p_{\rm T}$, then we have:
\begin{eqnarray}
& &\text{Ratio}(\rho^0/\pi^0)=\frac{d\sigma_{ \eta}}{dp_{T}}/\frac{d\sigma_{ \pi^0}}{dp_{T}} \nonumber \\
 &\approx&
\frac{\int F_{q}(\frac{p_{T}}{z_{h}})\ D_{q\to \rho^0}(z_{h}, p_{T})\frac{dz_{h}}{z_{h}^2}}
{\int F_{q}(\frac{p_{T}}{z_{h}})\ D_{q\to  \pi^{0}}(z_{h}, p_{T})\frac{dz_{h}}{z_{h}^2}}   \approx \frac{\Sigma_{q} D_{q\to \rho^0}(\left<z_{h}\right>, p_{T})}
{\Sigma_q D_{q\to  \pi^{0}}(\left<z_{h}\right>, p_{T}) }   \, .\nonumber
\end{eqnarray}
Therefore, while quark and gluon may lose different fractions of their energies, at very high $p_{\rm T}$ region, the ratio $\rho^0/\pi^0$ in $\rm A+A$ collisions should approximately be determined only by quark FFs in vacuum with the $p_{\rm T}$ shift because of the parton energy loss. As we can see in Fig.~\ref{fig:fragfunc}, the quark FFs at large scale $Q$ ($=p_{\rm T}$) change slowly with the variation of both $z_h$ and $p_{\rm T}$, then the ratio of $\rho^0/\pi^0$ in both $\rm A+A$ and $\rm p+p$ may approach to each other at larger $p_{\rm T}$. It is just as we have observed in the case for the yield ratio of $\eta/\pi^0$~\cite{Dai:2015dxa}.

{\bf Acknowledgments:} This research is supported by the MOST in China under Project No. 2014CB845404, NSFC of China with Project Nos. 11435004, 11322546, 11521064, and partly supported by the Fundamental Research Funds for the Central Universities, China University of Geosciences (Wuhan) (No. 162301182691)


\vspace*{-.6cm}

\begin{thebibliography}{99}

\bibitem{Wang:1991xy}
  X.~N.~Wang and M.~Gyulassy,
  Phys.\ Rev.\ Lett.\  {\bf 68}, 1480 (1992).

\bibitem{Gyulassy:2003mc}
  M.~Gyulassy, I.~Vitev, X.~N.~Wang and B.~W.~Zhang,
  In *Hwa, R.C. (ed.) et al.: Quark gluon plasma* 123-191
  [nucl-th/0302077].

\bibitem{Aamodt:2011vg}
  K.~Aamodt {\it et al.}  [ALICE Collaboration],
  Phys.\ Rev.\ Lett.\  {\bf 108}, 092301 (2012)
  [arXiv:1110.0121 [nucl-ex]].

\bibitem{Adler:2002tq}
  C.~Adler {\it et al.}  [STAR Collaboration],
  Phys.\ Rev.\ Lett.\  {\bf 90}, 082302 (2003)
  [nucl-ex/0210033].

\bibitem{Adare:2009vd}
  A.~Adare {\it et al.}  [PHENIX Collaboration],
  Phys.\ Rev.\ C {\bf 80}, 024908 (2009)
  [arXiv:0903.3399 [nucl-ex]].

\bibitem{Abelev:2009gu}
  B.~I.~Abelev {\it et al.}  [STAR Collaboration],
  Phys.\ Rev.\ C {\bf 82}, 034909 (2010)
  [arXiv:0912.1871 [nucl-ex]].




\bibitem{Vitev:2008rz}
  I.~Vitev, S.~Wicks and B.~W.~Zhang,
  JHEP {\bf 0811}, 093 (2008)
  [arXiv:0810.2807 [hep-ph]].

\bibitem{Vitev:2009rd}
  I.~Vitev and B.~W.~Zhang,
  Phys.\ Rev.\ Lett.\  {\bf 104}, 132001 (2010)
  [arXiv:0910.1090 [hep-ph]].

\bibitem{Dai:2012am}
  W.~Dai, I.~Vitev and B.~W.~Zhang,
  Phys.\ Rev.\ Lett.\  {\bf 110}, no. 14, 142001 (2013)
  [arXiv:1207.5177 [hep-ph]].

\bibitem{Aad:2010bu}
  G.~Aad {\it et al.}  [ATLAS Collaboration],
  Phys.\ Rev.\ Lett.\  {\bf 105}, 252303 (2010)
  [arXiv:1011.6182 [hep-ex]].

\bibitem{Chatrchyan:2011sx}
  S.~Chatrchyan {\it et al.}  [CMS Collaboration],
  Phys.\ Rev.\ C {\bf 84}, 024906 (2011)
  [arXiv:1102.1957 [nucl-ex]].

\bibitem{Kang:2014xsa}
  Z.~B.~Kang, R.~Lashof-Regas, G.~Ovanesyan, P.~Saad and I.~Vitev,
  Phys.\ Rev.\ Lett.\  {\bf 114}, no. 9, 092002 (2015)
  [arXiv:1405.2612 [hep-ph]].


\bibitem{Burke:2013yra}
  K.~M.~Burke {\it et al.}  [JET Collaboration],
  Phys.\ Rev.\ C {\bf 90}, no. 1, 014909 (2014)
  [arXiv:1312.5003 [nucl-th]];
  Z.~Q.~Liu, H.~Zhang, B.~W.~Zhang and E.~Wang,
  Eur.\ Phys.\ J.\ C {\bf 76}, no. 1, 20 (2016)
  [arXiv:1506.02840 [nucl-th]].

\bibitem{Chen:2010te}
  X.~F.~Chen, C.~Greiner, E.~Wang, X.~N.~Wang and Z.~Xu,
  Phys.\ Rev.\ C {\bf 81}, 064908 (2010)
  [arXiv:1002.1165 [nucl-th]].

\bibitem{Chen:2011vt}
  X.~F.~Chen, T.~Hirano, E.~Wang, X.~N.~Wang and H.~Zhang,
  Phys.\ Rev.\ C {\bf 84}, 034902 (2011)
  [arXiv:1102.5614 [nucl-th]].

\bibitem{Dai:2015dxa}
  W.~Dai, X.~F.~Chen, B.~W.~Zhang and E.~Wang,
  Phys.\ Lett.\ B {\bf 750}, 390 (2015)
  [arXiv:1506.00838 [nucl-th]].

\bibitem{Dai:2016zjy}
  W.~Dai and B.~W.~Zhang,
  arXiv:1612.05848 [hep-ph].

 


\bibitem{Agakishiev:2011dc}
  G.~Agakishiev {\it et al.}  [STAR Collaboration],
  Phys.\ Rev.\ Lett.\  {\bf 108}, 072302 (2012)
  [arXiv:1110.0579 [nucl-ex]].

\bibitem{Adare:2010pt}
  A.~Adare {\it et al.}  [PHENIX Collaboration],
  Phys.\ Rev.\ C {\bf 83}, 024909 (2011)
  [arXiv:1004.3532 [nucl-ex]].

\bibitem{Bala:2016hlf}
  R.~Bala, I.~Bautista, J.~Bielcikova and A.~Ortiz,
  Int.\ J.\ Mod.\ Phys.\ E {\bf 25}, no. 07, 1642006 (2016)
  [arXiv:1605.03939 [hep-ex]].

\bibitem{Liu:2006sf}
  W.~Liu, C.~M.~Ko and B.~W.~Zhang,
  Phys.\ Rev.\ C {\bf 75}, 051901 (2007)
  [nucl-th/0607047].



\bibitem{Brodsky:2008qp}
  S.~J.~Brodsky and A.~Sickles,
  Phys.\ Lett.\ B {\bf 668}, 111 (2008)
  [arXiv:0804.4608 [hep-ph]].

\bibitem{Chen:2008vha}
  X.~Chen, H.~Zhang, B.~W.~Zhang and E.~Wang,
  J.\ Phys.\  {\bf 37}, 015004 (2010)
  [arXiv:0806.0556 [hep-ph]].

\bibitem{Saveetha:2013jda}
  H.~Saveetha, D.~Indumathi and S.~Mitra,
  Int.\ J.\ Mod.\ Phys.\ A {\bf 29}, no. 07, 1450049 (2014)
  [arXiv:1309.2134 [hep-ph]].

\bibitem{Indumathi:2011vn}
  D.~Indumathi and H.~Saveetha,
  Int.\ J.\ Mod.\ Phys.\ A {\bf 27}, 1250103 (2012)
  [arXiv:1102.5594 [hep-ph]].

\bibitem{Hirai:2011si}
  M.~Hirai and S.~Kumano,
  Comput.\ Phys.\ Commun.\  {\bf 183}, 1002 (2012)
  [arXiv:1106.1553 [hep-ph]].



\bibitem{Guo:2000nz}
  X.~f.~Guo and X.~N.~Wang,
  Phys.\ Rev.\ Lett.\  {\bf 85}, 3591 (2000)
  [hep-ph/0005044].

\bibitem{Zhang:2003yn}
  B.~W.~Zhang and X.~N.~Wang,
  Nucl.\ Phys.\  A {\bf 720}, 429 (2003)
  [arXiv:hep-ph/0301195].

\bibitem{Zhang:2003wk}
  B.~W.~Zhang, E.~Wang and X.~N.~Wang,
  Phys.\ Rev.\ Lett.\  {\bf 93}, 072301 (2004)
  [nucl-th/0309040];
  A.~Schafer, X.~N.~Wang and B.~W.~Zhang,
  Nucl.\ Phys.\ A {\bf 793}, 128 (2007)
  [arXiv:0704.0106 [hep-ph]].


\bibitem{Kidonakis:2000gi}
  N.~Kidonakis and J.~F.~Owens,
  Phys.\ Rev.\ D {\bf 63}, 054019 (2001)
  doi:10.1103/PhysRevD.63.054019
  [hep-ph/0007268].



\bibitem{Lai:1999wy}
  H.~L.~Lai {\it et al.}  [CTEQ Collaboration],
  Eur.\ Phys.\ J.\ C {\bf 12}, 375 (2000)
  [hep-ph/9903282].

\bibitem{Dai:2017piq}
  W.~Dai, B.~W.~Zhang, H.~Z.~Zhang, E.~Wang and X.~F.~Chen,
  Eur.\ Phys.\ J.\ C {\bf 77}, no. 8, 571 (2017)
  [arXiv:1702.01614 [nucl-th]].


\bibitem{Hirano2001}
 T.~Hirano,
 Phys.\ Rev.\ C {\bf 65}, 011901 (2002).

\bibitem{HT2002}
 T.~Hirano and K.~Tsuda,
 Phys.\ Rev.\ C {\bf 66}, 054905 (2002).



\bibitem{Eskola:2009uj}
  K.~J.~Eskola, H.~Paukkunen and C.~A.~Salgado,
  JHEP {\bf 0904}, 065 (2009)
  [arXiv:0902.4154 [hep-ph]].

\bibitem{Adler:2003qi}
  S.~S.~Adler {\it et al.} [PHENIX Collaboration],
  Phys.\ Rev.\ Lett.\  {\bf 91}, 072301 (2003)
  doi:10.1103/PhysRevLett.91.072301
  [nucl-ex/0304022].



\bibitem{Aamodt:2010jd}
  K.~Aamodt {\it et al.} [ALICE Collaboration],
  Phys.\ Lett.\ B {\bf 696}, 30 (2011)
  doi:10.1016/j.physletb.2010.12.020
  [arXiv:1012.1004 [nucl-ex]].
  %
  %
\bibitem{Albino:2008fy}
  S.~Albino, B.~A.~Kniehl and G.~Kramer,
  Nucl.\ Phys.\ B {\bf 803}, 42 (2008)
  doi:10.1016/j.nuclphysb.2008.05.017
  [arXiv:0803.2768 [hep-ph]].

\bibitem{dEnterria:2013sgr}
  D.~d'Enterria, K.~J.~Eskola, I.~Helenius and H.~Paukkunen,
  Nucl.\ Phys.\ B {\bf 883}, 615 (2014)
  doi:10.1016/j.nuclphysb.2014.04.006
  [arXiv:1311.1415 [hep-ph]].

\bibitem{deFlorian:2014xna}
  D.~de Florian, R.~Sassot, M.~Epele, R.~J.~Hern��ndez-Pinto and M.~Stratmann,
  Phys.\ Rev.\ D {\bf 91}, no. 1, 014035 (2015)
  doi:10.1103/PhysRevD.91.014035
  [arXiv:1410.6027 [hep-ph]].
\end{thebibliography}
\end{document}